\documentclass[a4paper,epj]{svjour}

\usepackage{graphicx}
\usepackage{fancyhdr}
\usepackage{cite,mcite}
\usepackage[small]{caption}
\usepackage{amssymb}
\def\makeheadbox{{
 }}

\def\mytitle{My title} 
\def\myauthors{My name}  
\def\mytype{My type of session}
\def\mysession{My session}

\def\mytitle{Search for Charged Higgs Bosons at the LHC}
\def\myauthors{Martin Flechl (for ATLAS and CMS)}
\def\mytype{Contributed Talk}    
\def\mysession{Colliders - Higgs Phenomenology}

\begin{document}
\title{Search for Charged Higgs Bosons at the LHC}
\author{Martin Flechl
\thanks{\emph{Email:} martin.flechl@tsl.uu.se} (for the ATLAS and CMS collaborations)
}                     
%
%
\institute{Institutionen f\"{o}r K\"{a}rn- och Partikelfysik, Uppsala Universitet}
%
\date{}
\abstract{
The discovery potential for charged Higgs bosons at the ATLAS and CMS experiments is discussed with focus on the MSSM. Studies for several decay channels for the charged Higgs boson are presented both for production in top quark decays and gluon-gluon/gluon-bottom fusion.
} 
\maketitle
%

\section{Introduction}
\label{sec:intro}
Charged Higgs bosons ($H^\pm$) are predicted in several extensions of the Standard Model (SM). The simplest extension containing $H^\pm$ can be obtained by adding a second Higgs doublet to the SM, described by the Two Higgs Doublet Model (THDM). This model is particularly attractive since it describes the minimal required Higgs sector of supersymmetric models and is thus part of the Minimal Supersymmetric Standard Model (MSSM). Most simulation studies for $H^\pm$ searches at the LHC focus on the generic THDM and on the MSSM. The following results are for the MSSM ($m_h$-max scenario~\cite{carena02}) unless specified otherwise.

The THDM gives rise to five physical Higgs bosons, three neutral ones (h, H, A)  and a charged pair ($H^\pm$)\footnote{in the following, only $H^+$ is denoted but $H^-$ is always implicitely included}. At tree-level the THDM is fully described by two parameters which can be chosen to be the charged Higgs boson mass ($m_{H^+}$) and the ratio of the Higgs doublet vacuum expectation values ($\tan \beta$).
\begin{figure}[ht]
\begin{center}
\includegraphics[width=0.36\textwidth,height=4.3cm,angle=0]{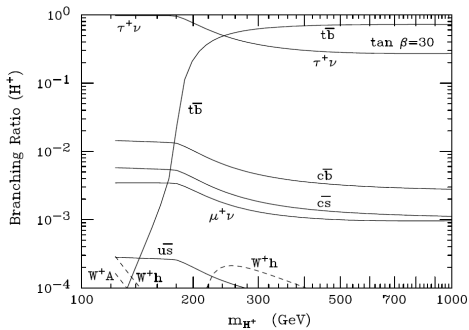}
\caption{$H^+$ SM branching ratios for $\tan\beta=30$ in an $m_h$-max-like MSSM scenario~\protect\cite{carena03}.}
\label{fig:br}
\end{center}
\end{figure}

The main $H^+$ production mode at the LHC are top quark decays ($t \to H^+b$) if the charged Higgs boson mass $m_{H^+} < m_t$ (the top quark mass), and gluon-gluon and gluon-bottom fusion ($gg \to tbH^+$, $gb \to tH^+$) if $m_{H^+} > m_t$. The charged Higgs boson couplings are proportional to the particle masses, and thus the dominant decay modes are $H^+ \to \tau\nu$ if $m_{H^+} < m_t$, and additionally $H^+ \to tb$ if $m_{H^+} > m_t$. 

In the MSSM, the lower experimental bound on the charged Higgs boson mass is about 80 GeV, independently of $\tan \beta$~\cite{lep01}. In the following sections, the charged Higgs boson discovery potential for the LHC experiments ATLAS and CMS estimated in simulation studies will be presented. The main decay channels will be discussed in Section~\ref{sec:top} for $m_{H^+} < m_t$ and in Section~\ref{sec:heavy} for $m_{H^+} > m_t$ while Section~\ref{sec:exotic} briefly discusses more exotic charged Higgs boson decay modes.

\section{Search for $H^+$ in Top Quark Decays}
\label{sec:top}
\begin{figure}[ht]
\begin{center}
\includegraphics[width=0.37\textwidth,angle=0]{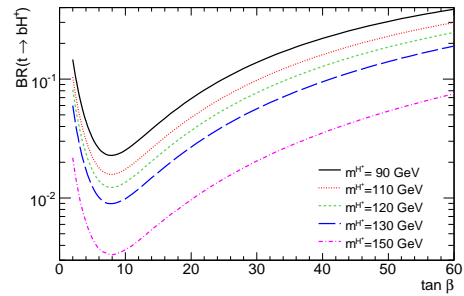}
\caption{Branching ratio $t \to bH^+$ as a function of $\tan \beta$ for different charged Higgs boson masses, $m_t=175$ GeV, with Pythia 6.3~\protect\cite{pythia}.}
\label{fig:xsecLow}
\end{center}
\end{figure}
The main production mode for $H^+$ at the LHC for $m_{H^+} < m_t$ is the decay of a top quark $t \to bH^+$ in $t\bar t$ events since the $t\bar t$ cross section is large ($794\;\textrm{pb}$ at NLO~\cite{kidonakis03}). The production of about 8 million top quark pairs per year at the LHC at low luminosity (${\cal L}=~\!\!10\;\mathrm{fb}^{-1}$ per year) is anticipated and with a typical expected branching ratio $t \to bH^+$ of a few percent this implies a sizeable $H^+$ production. The branching ratio $t \to bH^+$ as a function of $\tan \beta$ for different values of $m_{H^+}$ is shown in Figure~\ref{fig:xsecLow}.

Below the top quark mass, the charged Higgs boson decays almost exclusively to a tau lepton and a neutrino (except for very low $\tan \beta$ values). Recent searches have focused on hadronic $\tau$ decays of the $H^+$ while both hadronic and semileptonic decays of the second top quark in the event are considered.

\subsection{$t\bar{t} \to H^+bWb$ with $H^+ \to \tau\nu$, $W \to l\nu$}
\label{sec:top_Whad}
This channel was investigated by Baarmand et~al.~\cite{baarmand06} for the CMS experiment. The analysis is a counting experiment since $m_{H^+}$ cannot be reconstructed due to the presence of neutrinos from two sources. The study was carried out using full detector simulation including pile-up. As main backgrounds, $t\bar{t}$ with one $t \to b\ell\nu$ and $W+3$ jets with $W \to \ell\nu$ were considered. A sequence of selection cuts based on the event signature was applied to events accepted by a lepton trigger. At least 3 jets with $E_T > 40$ GeV were required, exactly one of them being b-tagged. A cut on at least one $\tau$ jet with $E_T > 40$ GeV and a leading track with a momentum of at least 0.8 of the $\tau$ jet transverse energy is applied next. The sum of the charge of the $\tau$ jet and the lepton are required to be zero. Finally a cut on the missing transverse energy of 70 GeV is applied.
\begin{figure}[ht]
\begin{center}
\includegraphics[width=0.36\textwidth,angle=0]{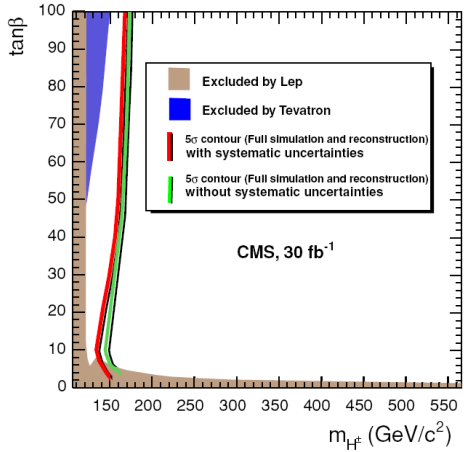}
\caption{$H^+$ discovery contour for $t\bar{t} \to H^+bWb$ with $H^+ \to \tau\nu$, $W \to l\nu$ at CMS~\protect\cite{baarmand06}.}
\label{fig:discBaar}
\end{center}
\end{figure}

The resulting discovery contour including systematic uncertainties is shown in Figure \ref{fig:discBaar}. A discovery potential for all values of $\tan \beta$ is given for $m_{H^+} \lesssim 135$ GeV and an integrated luminosity of $30\;\textrm{fb}^{-1}$ (three years at low luminosity) and the reach goes up to about $m_{H^+} = 170$ GeV for very high values of $\tan \beta$.

\subsection{$t\bar{t} \to H^+bWb$ with $H^+ \to \tau\nu$, $W \to qq$}
\label{sec:top_Wlep}
\begin{figure}[ht]
\begin{center}
\includegraphics[width=0.32\textwidth,angle=0]{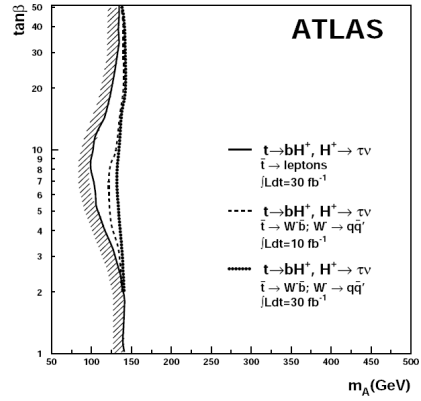}
\caption{$H^+$ discovery contour for $t\bar{t} \to H^+bWb$ with $H^+ \to \tau\nu$, $W \to qq$ at ATLAS~\protect\cite{biscarat03}}
\label{fig:discBisc}
\end{center}
\end{figure}
This analysis has been performed by Biscarat et al.~\cite{biscarat03} for the ATLAS experiment. It is based on a parametrized simulation of the ATLAS detector. The considered backgrounds are $t\bar t$ and QCD. The trigger condition is either a $\tau$ jet or a jet, combined with missing transverse energy. In the first stage of the event selection chain, requirements following closely the event signature are applied: a $\tau$ jet, exactly two b jets, at least two light jets above certain $p_T$ thresholds, a veto on isolated leptons, $E_T^{miss} > 45$ GeV and the reconstruction of the $W$ and its associated top quark within mass windows. Additionally, a set of cuts aimed at the $t\bar t$ pattern is applied to reduce the QCD background, consisting of angular and momentum relations of the top quarks and their decay products. Eventually cuts on the relation of the $\tau$ and the $b$ on the $H^+$ side are applied, as well as on the $H^+$ transverse mass (only for the analysis of $m_{H^+} > 113$ GeV).

The resulting discovery contour is shown in Figure~\ref{fig:discBisc}. For $30\;\textrm{fb}^{-1}$, a discovery potential is given for all $\tan \beta$ for $m_A \lesssim 135$ GeV, corresponding to $m_{H^+} \approx 155$ GeV. Systematic uncertainties are included.

\section{Search for $H^+$ in gg- and gb-fusion}
\label{sec:heavy}
\begin{figure}[ht]
\begin{center}
\includegraphics[width=0.4\textwidth,angle=0]{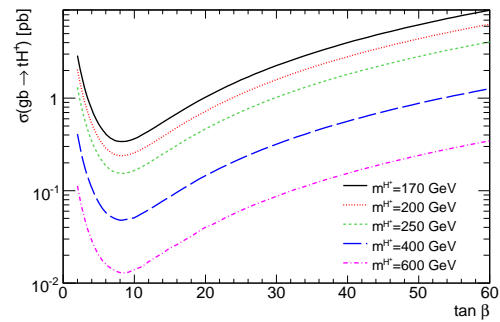}
\caption{NLO cross section $gb\to tH^+$ as function of $\tan \beta$ for various charged Higgs boson masses, $m_t=175$ GeV~\protect\cite{plehn02,boos03}}
\label{fig:xsecHigh}
\end{center}
\end{figure}
At the LHC, charged Higgs bosons with $m_{H^+}\!>~m_t$ would be dominantly produced in the processes $gg \to tbH^+$ and $gb \to tH^+$. These so-called ``twin processes'' correspond to two different approximations of the same process which leads to an overlap in certain regions of phase space. This overlap is resolved by the event generator MATCHIG~\cite{alwall04} which is used in recent and ongoing ATLAS studies. The $H^+$ cross section in $gb$/$gg$-fusion is much smaller than in $t\bar{t}$ decays and decreases rapidly with mass, as shown in Figure \ref{fig:xsecHigh}.

The dominating decay modes are $H^+ \to \tau\nu$ and $H^+ \to tb$ , with the branching ratio of the latter being 1-2 orders of magnitude higher for low $\tan \beta$ and/or high $m_{H^+}$. However, the $\tau$ channel provides a much cleaner signature, as will be shown in the following sections.

\subsection{$gg/gb \to t[b]H^+$ with $H^+ \to tb$}
\label{sec:heavy_tb}
This channel with the final state $b\bar{b}b[\bar{b}]\ell\nu q\bar{q}$ was investigated by Lowette et~al.~\cite{lowette06} for CMS and by Assamagan et~al.~\cite{assamagan04} for ATLAS. In the following the focus will be on the CMS study (differences to the ATLAS study will be pointed out at the end) which was carried out using full detector simulation including pile-up. The main backgrounds were considered to be $t\bar{t}+b$, $t\bar{t}+b\bar{b}$ and $t\bar{t}+$jet(s). For the analysis requiring 4 $b$-tags, only $gg \to tbH^+$ was simulated; an additional 3 $b$-tag analysis with parametrized detector simulation used $gb \to tH^+$ events. In both cases the cross section was rescaled to the NLO value for the $tH^+X$ final state. Only final states with $\ell = \mu$ were considered.
\begin{figure}[ht]
\begin{center}
\includegraphics[width=0.24\textwidth,angle=0]{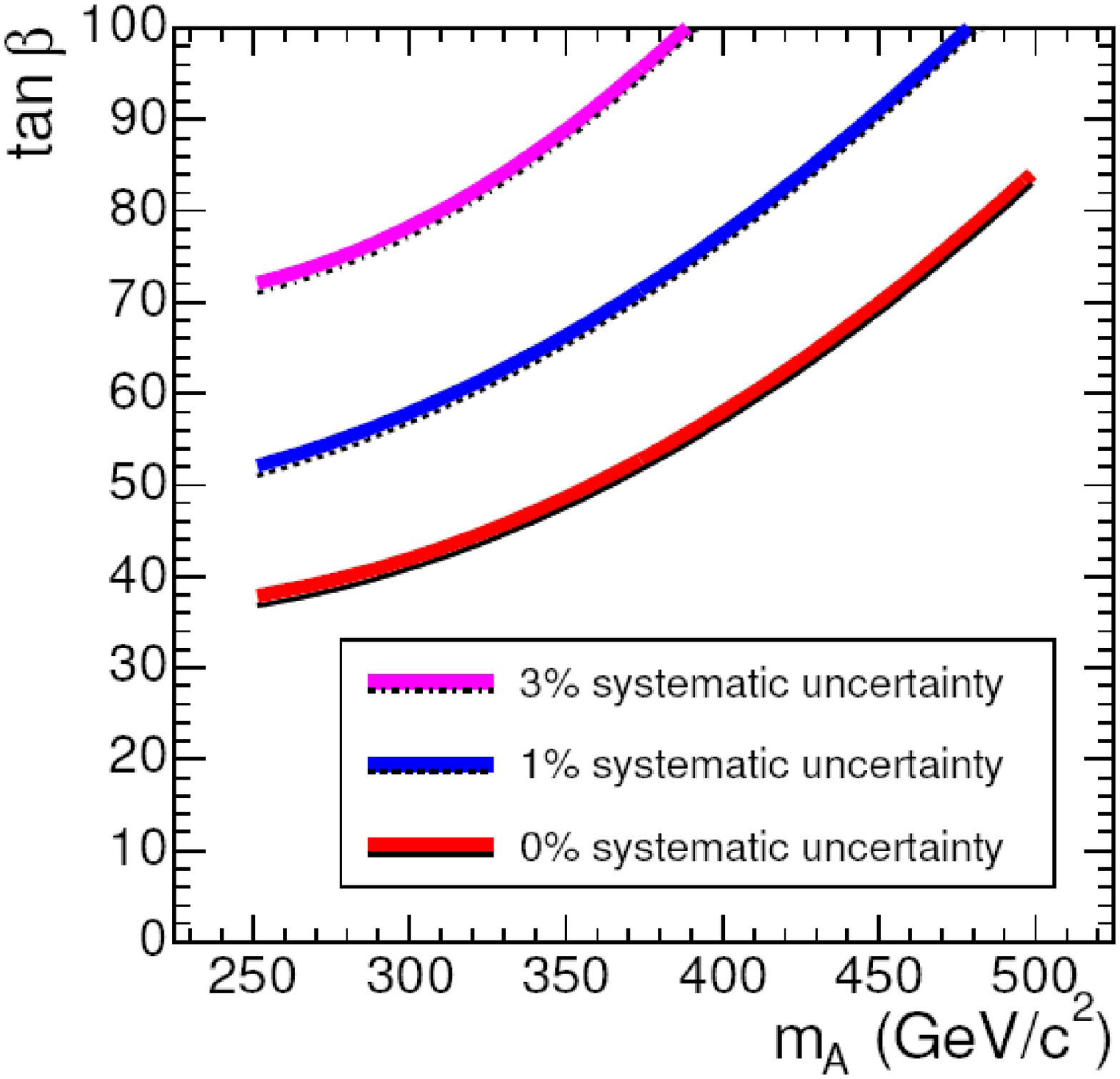}
\includegraphics[width=0.24\textwidth,angle=0]{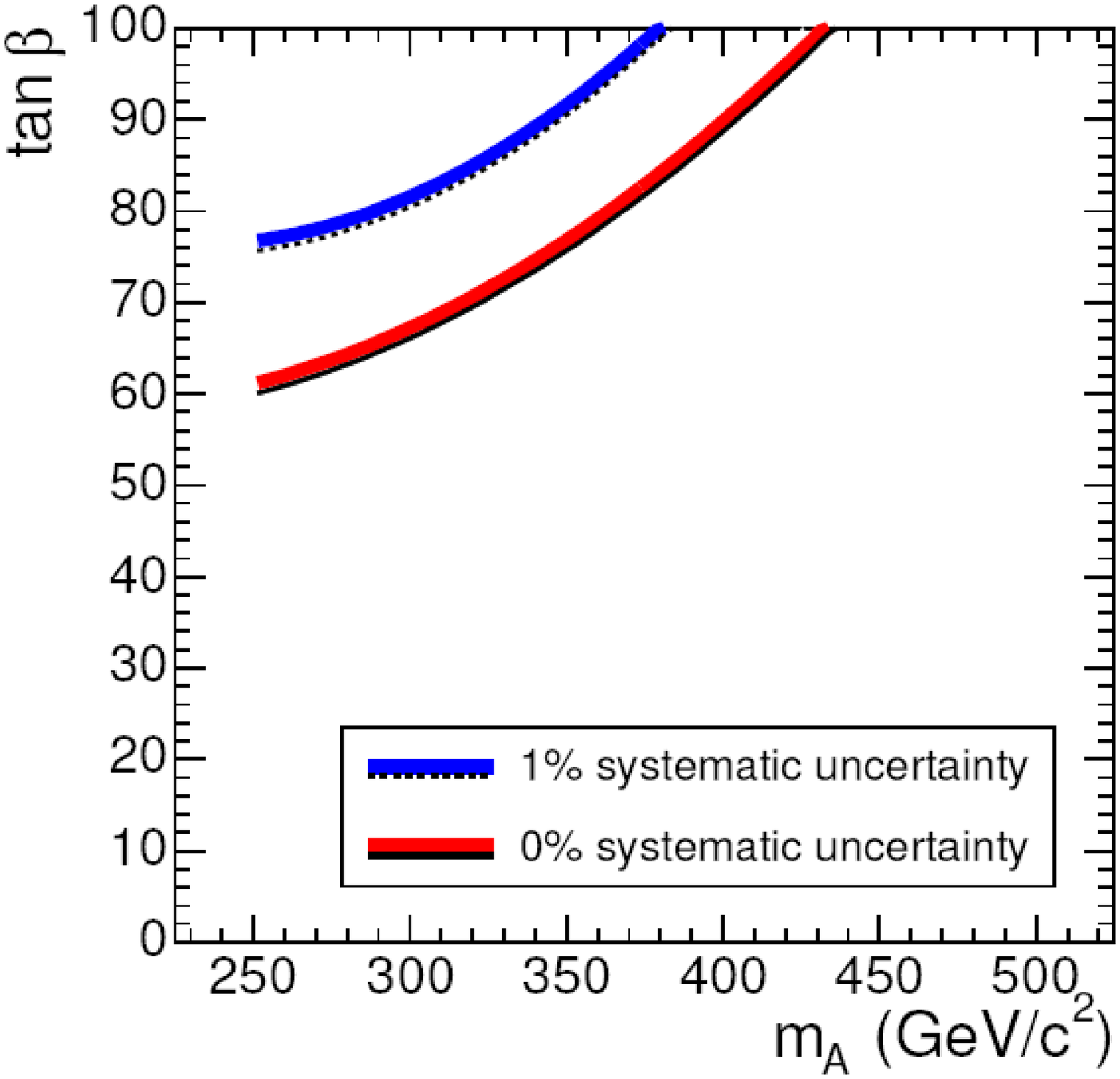}
\caption{CMS discovery contour for $gg/gb \to t[b]H^+$ with $H^+ \to tb$~\protect\cite{lowette06}. Left: 3 $b$-tag analysis. Right: 4 $b$-tag analysis.}
\label{fig:discLow}
\end{center}
\end{figure}
The following requirements were placed on events accepted by a muon trigger for the 4 $b$-tag analysis (in brackets: 3 $b$-tag analysis): a reconstructed muon with $p_T > 20$ GeV, at least 6 (5) jets with $p_T > 25$ GeV, 4 (3) of them $b$-tagged, and a kinematic fit imposing mass constraints for both W bosons and top quarks. In the next step a likelihood method is used to suppress the combinatorial background, using information from the kinematic fit and properties of the $b$ jets. Finally the $t\bar{t}+X$ background is further suppressed via a selection likelihood using $b$-tagging information (in the 3 $b$-tag analysis, additionally information about the light jets and from the kinematic fit is used).\\
The main difference between this study and the corresponding ATLAS analysis~\cite{assamagan04} is that the latter is based upon parametrized detector simulation without pile-up, that it considers the electron mode in addition to the muon mode and that only $gg \to tbH^+$ requiring 4 $b$-tags is investigated. Additionally, different $b$ jet-related variables are used for background suppression. The resulting discovery contour for CMS is shown in Figure \ref{fig:discLow}, the results for ATLAS are similar. As can be seen, even assuming very low systematic uncertainties, very little sensitivity is given in the MSSM parameter space.

\subsection{$gg/gb \to t[b]H^+$ with $H^+ \to \tau\nu$}
\label{sec:heavy_taunu}
The $\tau$ decay mode for a heavy $H^+$ was studied for ATLAS by Mohn et~al.~\cite{mohn07}, and for CMS by Kinnunen~\cite{kinnunen06}. In the ATLAS study, the considered backgrounds are $t\bar{t}$, single top, QCD dijets and $W$+jets. Full detector simulation was used for the $H^+$ signal and a parametrized detector response for the backgrounds. The event generator MATCHIG~\cite{alwall04} allowed for the first time a consistent treatment of the transition region $m_{H^+} \approx m_t$. The selection cuts were optimised for three different mass ranges ($m_{H^+} \leq 250$ GeV, $250<m_{H^+}<450$ GeV and $m_{H^+} \geq 450$ GeV). The following selection cuts are applied (for the low / medium / high mass range): one $\tau$ jet with $p_T > 65 / 80 / 100$ GeV within $|\eta| < 1.1 / 1.2 / 1.3$, $E_T^{\textsuperscript{miss}} > 120 / 135 / 165$ GeV, no isolated lepton, at least 3 jets (exactly one of them $b$-tagged), the W boson and top quark masses reconstructed within 25 GeV-windows, and  $p_T^\tau / p_T^{\textsuperscript{add. jet}} > 6.0 / 5.5 / 5.0$ (``add. jet'' is the hardest jet not used for the top quark reconstruction).\\
The main differences in the CMS study are that full detector simulation with pile-up and trigger simulation was used both for signal and background. No matched signal event generation has been performed, only $gg \to tbH^+$ was simulated and scaled to the matched production cross section. Instead of the cut on $p_T^\tau / p_T^{\textsuperscript{add. jet}}$, a veto on additional central jets was applied. To exploit the $\tau$ helicity correlations, an additional cut on the ratio of the leading $\tau$ track momentum and the energy of the $\tau$ jet was applied.
\begin{figure}[htb]
\parbox[t]{0.24\textwidth}{
    \begin{center}
    \includegraphics[height=3.0cm,angle=0]{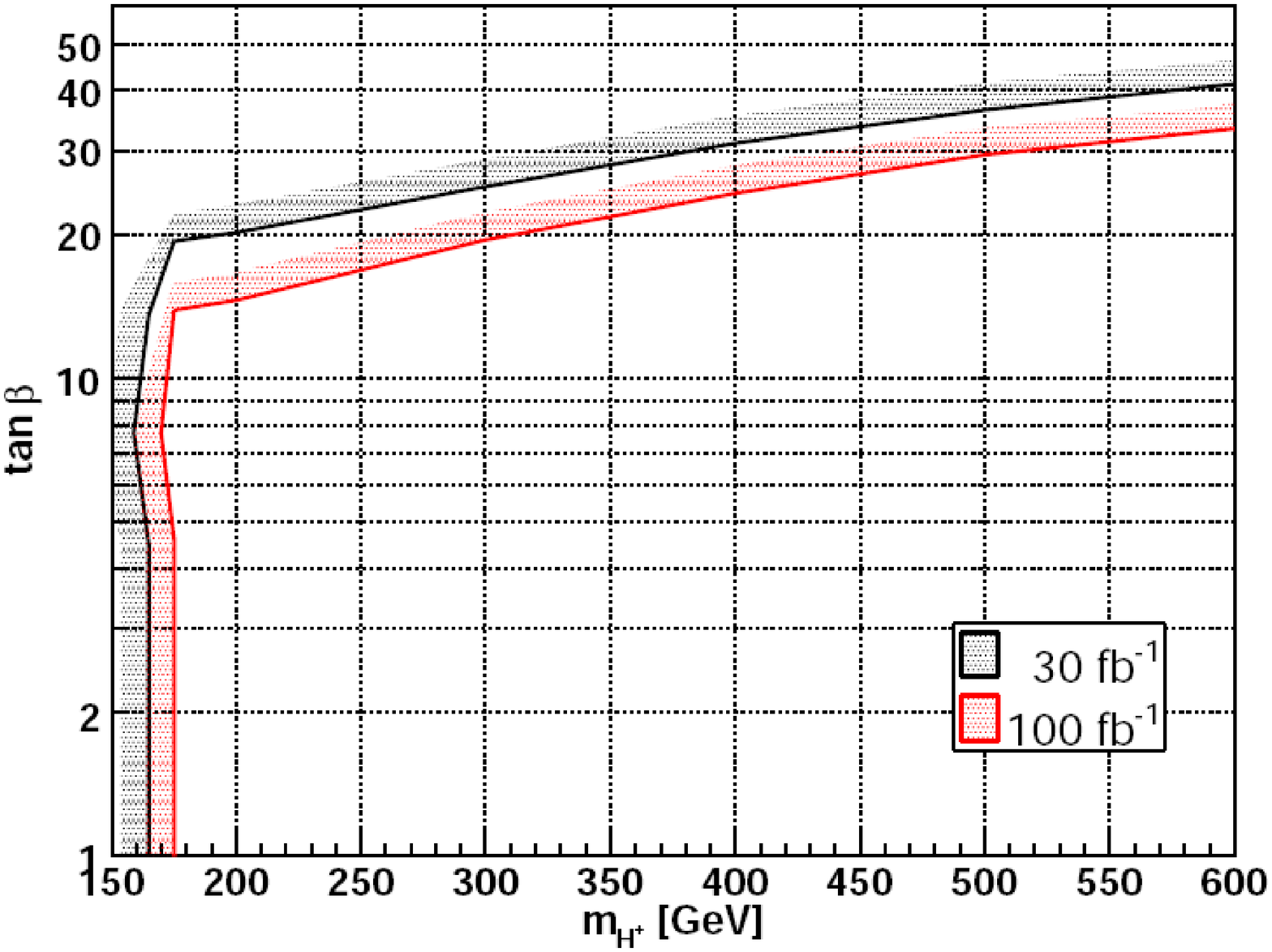}
    \captionsetup{width=3.5cm}
    \captionsetup{aboveskip=0cm}
    \caption{ATLAS discovery contour for heavy $H^+~\!\!\to~\!\!\tau\nu$~\protect\cite{mohn07} excluding systematic uncertainties.}
    \label{fig:discMohn}
    \end{center}
}
\parbox[t]{0.24\textwidth}{
    \begin{center}
    \includegraphics[height=3.0cm,angle=0]{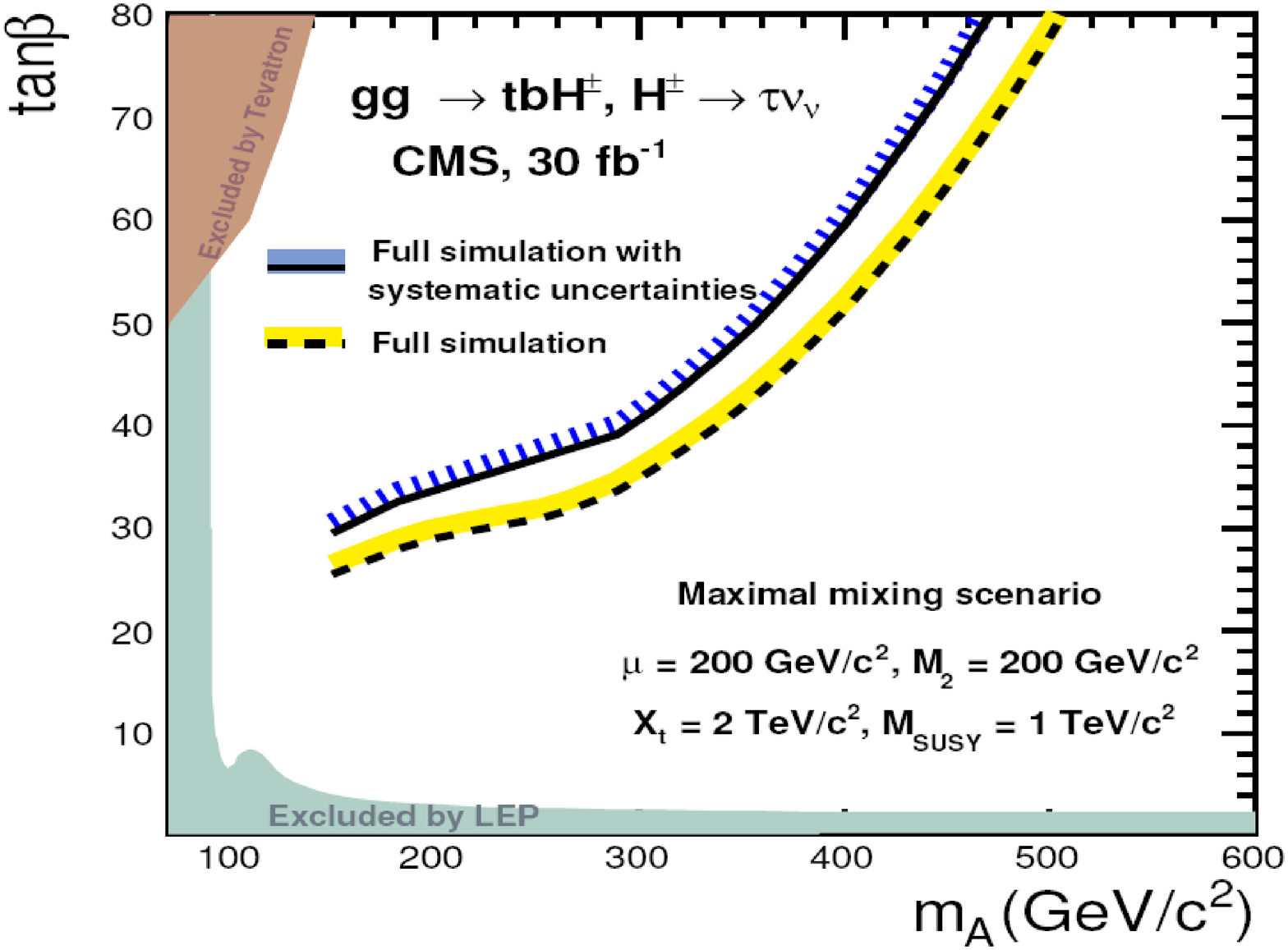}
    \captionsetup{width=3.5cm}
    \captionsetup{aboveskip=0cm}
    \caption{CMS discovery contour for heavy $H^+~\!\!\to~\!\!\tau\nu$~\protect\cite{kinnunen06}.}
    \label{fig:discKinn}
    \end{center}
}
\end{figure}
The discovery contours are shown in Figure \ref{fig:discMohn} for ATLAS, and Figure \ref{fig:discKinn} for CMS. The sensitivity is significantly better than in the $H^+ \to tb$ channel since the smaller signal-to-background ratio leads to more robustness with regard to systematic uncertainties. However, the region $\tan \beta \lesssim 20$ is currently uncovered for $m_{H^+}>m_t$ at an integrated luminosity of $30\;\textrm{fb}^{-1}$ and the discovery potential decreases rapidly with increasing $m_{H^+}$ due to the diminishing cross section.

\section{Search for $H^+$ in Exotic Channels}
\label{sec:exotic}
Charged Higgs boson simulation studies were also performed in non-SM and in rare $H^+$ channels. Two examples are briefly presented. 

The mode $H^+ \to Wh^0$ has been studied by Assamagan~\cite{assamagan99}, and $H^+ \to WH^0$ in a large mass splitting MSSM scenario by Mohn et al.~\cite{mohn03}. Both studies conclude that even in the most favourable scenarios, with knowledge of the neutral Higgs boson masses and at high luminosity no MSSM discovery sensitivity is given via these channels.

\begin{figure}[ht]
\begin{center}
\includegraphics[width=0.3\textwidth,angle=0]{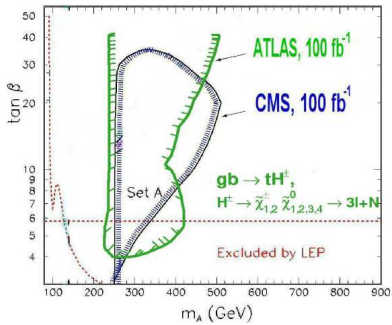}
\caption{$H^+ \to \tilde{\chi}^\pm_{1,2} \tilde{\chi}^0_{1,2,3,4}$ discovery contour for ATLAS and CMS for a favourable MSSM scenario~\protect\cite{hansen05}. Systematic uncertainties are not taken into account.}
\label{fig:discHan}
\end{center}
\end{figure}

The channel $H^+ \to \tilde{\chi}^\pm_{1,2} \tilde{\chi}^0_{1,2,3,4}$ for a heavy charged Higgs boson produced via $gb \to tH^+$ was studied by Hansen et~al.~\cite{hansen05} for ATLAS and Bisset et~al.~\cite{bisset03} for CMS using parametrized detector simulation. Both studies represent counting experiments and require the measurement of cross sections of several SUSY processes in order to be able to evaluate the background. Particularly high precision is needed in the case of the ATLAS study since the signal-to-background ratio is low (typically, $\approx 0.1$). Discovery contours for both studies for a very favourable MSSM scenario and without systematic uncertainties are shown in Figure~\ref{fig:discHan}. The most important conclusion is that in principle there is sensitivity in the intermediate $\tan \beta$-region not covered by any other channels.
\vspace{-1.0cm}

\section{Combined Discovery Contours}
\label{sec:comb}
\begin{figure}[htb]
\parbox[t]{0.24\textwidth}{
    \begin{center}
\includegraphics[width=0.25\textwidth,height=3.9cm,angle=0]{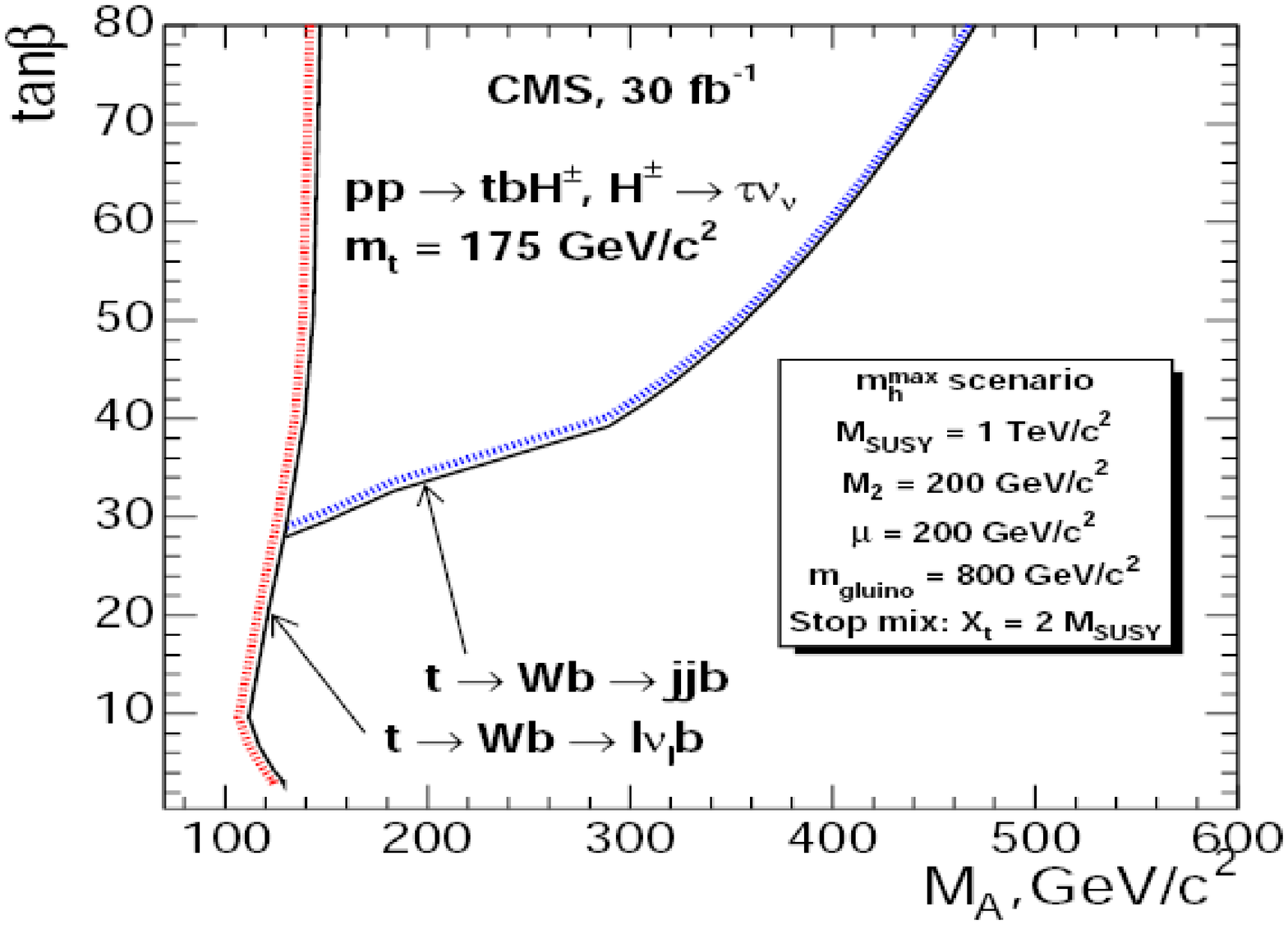} 
    \captionsetup{width=3.9cm}
    \captionsetup{aboveskip=0cm}
\caption{CMS combined $H^+$ discovery contour~\protect\cite{cmstdr06}.}
\label{fig:discCMS}
    \end{center}
}
\parbox[t]{0.24\textwidth}{
    \begin{center}
\includegraphics[width=0.236\textwidth,height=3.95cm,angle=0]{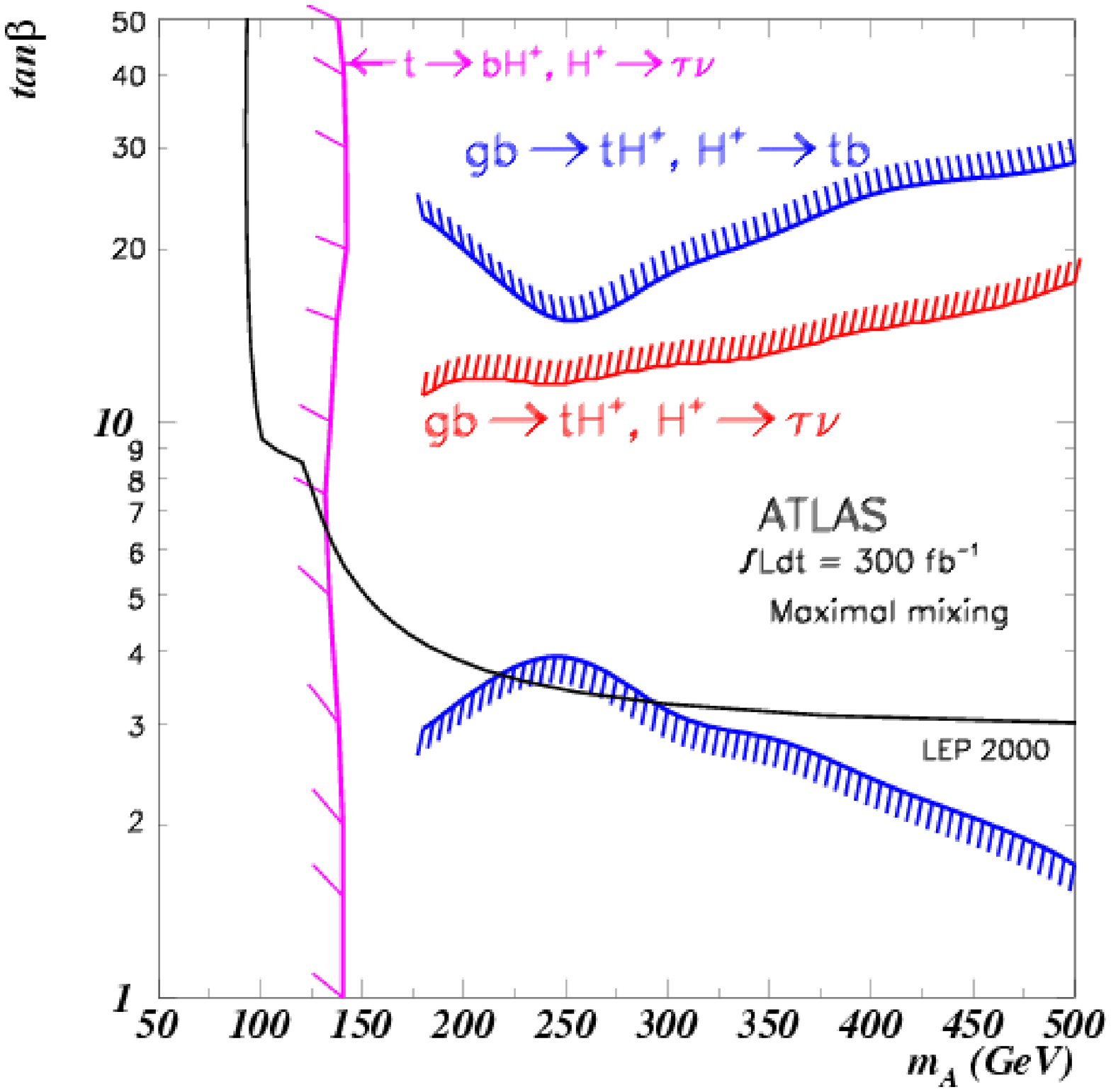} 
    \captionsetup{width=3.9cm}
    \captionsetup{aboveskip=0cm}
\caption{ATLAS combined $H^+$ discovery contour~\protect\cite{hansen05}.}
\label{fig:discATLAS}
    \end{center}
}
\end{figure}

The combined discovery contours are shown in Figure~\ref{fig:discCMS} (CMS) and Figure \ref{fig:discATLAS} (ATLAS). The two contours cannot be compared as the ATLAS contour is based on parametrized detector simulation, does not include systematics and assumes ${\cal L}=300\;\textrm{fb}^{-1}$ while the CMS contour is for ${\cal L}=30\;\textrm{fb}^{-1}$, and based on full detector simulation including pile-up and systematics. An ATLAS update is expected for the end of 2007.

The current studies suggest that the parameter space up to $m_{H^+}$ values close to the top quark mass will be covered by the LHC in a few years of running at low luminosity except for a small region of intermediate $\tan \beta$ ($\approx 4-15$) which requires high luminosity runs. For $m_{H^+} > m_t$, sensitivity is present only for $\tan \beta$ higher than about $15-20$, and the discovery potential diminishes rapidly with increasing $m_{H^+}$.

\section{Conclusions and Outlook}
\label{sec:conc}
$H^+ \to \tau\nu$ is the prime MSSM charged Higgs boson discovery channel at the LHC. $H^+ \to tb$ does not add any sensitivity. The MSSM parameter space will be covered for all $\tan \beta$ for values of $m_{H^+}$ up to close to $m_t$. For $m_{H^+} > m_t$, only $\tan \beta$ greater than about 15-20 will be covered in low luminosity runs (and the uncovered region increases with $m_{H^+}$). $H^+$ in the gap region $4 < \tan \beta < 10$, which will probably not even be covered in high luminosity runs by SM decay modes, could be discovered in decays to SUSY particles after measurement of SUSY cross sections.

All standard channels investigated with parametrized detector response are in the process of being updated with full detector simulation. Additionally, simulation results for new channels like $H^+ \to \tau\nu$ with leptonic $\tau$ decays as well as for associated production $AH^+$ are expected in the near future. Preparations for the first LHC data in 2008 are in their final phase knowing that charged Higgs bosons could be the first Beyond the Standard Model-signal we see.
\bibliographystyle{atlasstylem}
\bibliography{MartinFlechl_susy07}

\begin{mcbibliography}{10}

\bibitem{carena02}
Carena M S et al,
\newblock  Eur. Phys. J. {\bf C26}  (2003) 601--607\relax
\relax
\bibitem{carena03}
Carena M S et al,
\newblock  Prog. Part. Nucl. Phys. {\bf 50}  (2003) 63--152\relax
\relax
\bibitem{lep01}
LEP Higgs Working Group,
\newblock  hep-ex/0107031\relax
\relax
\bibitem{pythia}
Sj{\"o}strand T et al,
\newblock  JHEP {\bf 05}  (2006) 026\relax
\relax
\bibitem{kidonakis03}
Kidonakis N et al,
\newblock  Phys. Rev. {\bf D68}  (2003) 114014\relax
\relax
\bibitem{baarmand06}
Baarmand M et al,
\newblock  J. Phys. {\bf G32}  (2006) N21\relax
\relax
\bibitem{biscarat03}
Biscarat C et al,
\newblock  {ATL-PHYS-2003-038}\relax
\relax
\bibitem{plehn02}
Plehn T,
\newblock  Phys. Rev. {\bf D67}  (2003) 014018\relax
\relax
\bibitem{boos03}
Boos E et al,
\newblock  Phys. Rev. {\bf D69}  (2004) 094005\relax
\relax
\bibitem{alwall04}
Alwall J et al,
\newblock  JHEP {\bf 12}  (2004) 050\relax
\relax
\bibitem{lowette06}
Lowette S et al,
\newblock  {CERN-CMS-NOTE-2006-109}\relax
\relax
\bibitem{assamagan04}
Assamagan K et al,
\newblock  Eur. Phys. J. {\bf C39S2}  (2005) 25--40\relax
\relax
\bibitem{mohn07}
Mohn B et al,
\newblock  {ATL-PHYS-PUB-2007-006}\relax
\relax
\bibitem{kinnunen06}
Kinnunen R,
\newblock  {CERN-CMS-NOTE-2006-100}\relax
\relax
\bibitem{assamagan99}
Assamagan K,
\newblock  {ATL-PHYS-99-025}\relax
\relax
\bibitem{mohn03}
Mohn B et al,
\newblock  {ATL-PHYS-PUB-2005-017}\relax
\relax
\bibitem{hansen05}
Hansen C et al,
\newblock  Eur. Phys. J. {\bf C44S2}  (2005) 1--9\relax
\relax
\bibitem{bisset03}
Bisset M et al,
\newblock  Eur. Phys. J. {\bf C30}  (2003) 419--434\relax
\relax
\bibitem{cmstdr06}
Ball A et al,
\newblock  {CMS-TDR-008-2}, 2006\relax
\relax
\end{mcbibliography}





\end{document}